\documentclass[11pt,a4paper]{article}
\usepackage{geometry}  
\usepackage[british]{babel} % [british]
\usepackage[T1]{fontenc}
\usepackage[utf8]{inputenc}
\usepackage[pdftex]{graphicx} % or [dvips]
\usepackage{color}
\usepackage[hyphens]{url}
\usepackage{fourier,charter,booktabs} 
\usepackage{enumitem} \setlist[1]{itemsep=0.4ex} \setlist[2]{itemsep=0.2ex}
\usepackage{varioref}\labelformat{equation}{(#1)}
\usepackage[round]{natbib}
\usepackage{amsfonts}
\newcommand{\inv}{^{-1}}
\newcommand{\Real}{\mathbb{R}}
\newcommand{\Rlang}{{\normalfont\textsf{R}}{}} 

\newcommand{\half}{\mbox{$\textstyle \frac{1}{2}$}}   
\newcommand{\eps}{\varepsilon}
\title{\bf On the use of ordered factors as explanatory variables}
\author{{\Large Adelchi Azzalini} \\  \large 
  Dipartimento di Scienze Statistiche \\
  Università degli Studi di Padova, Italia
 }
\date{\small \today} 
\begin{document}
\maketitle
\vspace{1ex}
\textsl{ 
The present version of the paper essentially coincides with the one
appearing in \emph{Stat} {\bf 12}, e624 (2023), freely accessible at 
\texttt{\url{https://doi.org/10.1002/sta4.624}}. 
The only difference is in the very last paragraph of the main body of the text.
}
\vspace{1ex}
\begin{quote}
\textbf{Abstract} 
Consider a regression or some regression-type model for a certain response variable 
where the linear predictor includes an ordered factor among the explanatory variables. 
The inclusion of a factor of this type can take place is a few different ways,
discussed in the pertaining literature.
The present contribution proposes a different way of tackling this problem,
by constructing a numeric variable in an alternative way with respect to
the current methodology.
The proposed techniques appears to retain the data fitting capability of 
the existing methodology, but with a simpler interpretation of 
the model components.
\end{quote}
\vspace{1ex}

% \keywords{ordered factors, ordered categorical variables, explanatory variables, 
% factor coding, linear predictor,  transformations of the normal distributions}

\maketitle

%==============================================================================

\section{Context and aim}
The statistical analysis of data  arising from categorical variables, 
or factors, of ordered type involves specialized techniques.
A valuable general treatment of the pertaining methods is provided by 
\citet{agresti:2010}. 
To introduce notation, denote by $X$ the ordered factor under consideration, 
whose levels  $1,\dots, K$ are in increasing order in some recognized sense, 
but they do not represent values on a numerical scale.
For the study of relationship among a set of variables, the situation more
extensively discussed in the specialized literature refers to the case 
where $X$ represents the response variable.

The present note deals instead with a different situation,
namely with the case where $X$ plays the role of 
an explanatory variable, not the response one.
This other setting has received substantially less attention in the literature.
After a brief survey of the existing options for making use of $X$ 
as an explanatory variable in a linear model or a generalized
linear model, we shall put forward a new proposal.
 
The inclusion of an ordered factor in a linear predictor implies its 
transformation into some numeric variable, or possibly a number of them.
For the selection of a suitable numerical score scheme to be used
as an explanatory variable, 
\citet{armitage:1955} % cf p.378: `In the absence of any...`
and \citet{grau:korn:1987} % cf p.472 alla fine di \S 1.
provide essentially similar recommendations, as follows:
(i) the ideal route is to identify some numerical score 
with a convincing subject-matter interpretation;
(ii)~in the absence of any \emph{a priori} knowledge towards such a 
choice, one must pick-up some numeric values, and
equally spaced values are the natural option to consider,
at least as the initial choice.
Since route (ii) involves a subjective choice,
it has generated much debate over the years, and still it does,
as demonstrated for instance  by the views presented fairly 
recently by \citet{pasta:2009} and \citet{williams:2020}.

In the common case that one chooses to employ equally spaced values
as scores,  it is however appropriate to test 
the  assumption of  linearity of the effect. 
For instance, Section~2 of \citet{armitage:1955} presents a
decomposition of the variability due to $X$ into two components,
the one due to linearity and the one due to departure from linearity.

Moving further in this direction, one can introduce more than one
numeric variable. A specific option is to consider a sequence
of orthogonal polynomials evaluated at $(1,\dots, K)$ 
or at an affine transformation of this sequence, 
so that the total variability associated to 
$X$ can be decomposed into components associated to the linear, quadratic,  
and other polynomial terms, with components up to a maximal degree $K-1$. 
An early exemplification  of this procedure is provided by 
\citet[pp.\,70--77]{winer:1962}, in a numerical illustration involving 
the effect of a  six-level factor  representing `complexity' 
of a certain visual display, whose effect is decomposed into  
linear, quadratic,  cubic and residual components.

This logic is currently standard practice. 
Specifically, in the \Rlang\ computing environment \citep{R-lang:2022}, 
ordered factors are handled in this way unless an alternative choice is explicitly made;
see the documentation of command \texttt{contr.poly} for details.
This strategy is in line with the indications of \citet[p.\,32]{cham:hast:1993} 
who prescribed that
``ordered factors are coded so that individual coefficients represent orthogonal
polynomials if the levels of the factor were actually equally spaced numeric
values''. 

However, alternative directions are possible. 
Specifically, \cite{gert:tutz:2009} and \cite{tutz:gert:2016} have proposed
to consider regularization methods for the parameters based on 
``a penalty which enforces that estimates of coefficients for adjacent 
categories are not too far apart''. This is a  way of
approaching the problem which, as highlighted by the discussion in 
\citet[pp.\,347--8]{gert:tutz:2009}, follows a  different logic
from the alternative of assigning scores, as we do here.

While an adequate fit to the data remains a basic requirement for any
candidate formulation, the key feature of the present proposal is a simplification 
of the modelling process via a method which delivers a scoring system.
The construction of a scoring system appears to be a desirable feature 
for many applied researchers, since it typically simplifies interpretation;
this aspect will be illustrated in the practical situations presented  later.
This feature makes a qualitative difference with a number of other proposals,
such as those based on regularization methods recalled  in the previous
paragraph, but also with other formulations. 
In this view, a numerical comparison with alternative methods 
beyond the standard ones does not seem to be crucial.

For the sake of completeness, we mention the connection with the paper
by \citet{wins:mare:1984}. However, for ordinal independent variables,
they consider a somewhat different setting,
namely the case where ordered variables occur as 
``independent or intervening variables in structural equation models''.  

The rest of this note is dedicated to a alternative way of constructing
a set of scores, where the basic values $1,\dots, K$ are
replaced by a set of not equally-spaced values.
The method works by choosing the non-equally-spaced values in place of 
quadratic or higher-degree polynomial values, 
with the conceptual advantage of providing numeric scores 
which reflect the actual data behaviour. 
The overall aim is to construct  a numeric response variable 
which replaces the original  ordered factor, retaining a similar 
fitting ability but allowing  a simpler interpretation.
The procedure is illustrated by two numerical exemplifications,
followed by a set of concluding remarks.
%---------------------------------------------------------------------

\section{A method for choosing scores} \label{s:method}

%----------------
\subsection{The rationale of the formulation}   

The basic scores $1,\dots,K$ are equivalent to the scaled values
$u_k=k/(K+1)\in(0,1)$ for $k=1,\dots,K$.
In case the original scores span a different set of values, 
such as $0,1,\dots, K-1$, there exists some other affine
transformation to achieve the same $u_{k}$'s values;
the only requirement is that the original scores are equally spaced.
Our aim is to select a non-linear transformation of the $u_{k}$'s,
which can better fit the data pattern than the equally-spaced 
scores.
This plan can be  broadly viewed as a replacement of the above-mentioned
suggestion by \citet{armitage:1955} and \citet{grau:korn:1987} 
of identifying scores with subject-matter interpretation,
in the frequently occurring cases where a specification 
with such an ideal genesis is not feasible.

Take now into consideration two  simple facts: 
(a)~the $u_{k}$'s correspond to a set of equally-spaced quantiles 
of the $U(0,1)$ distribution,
and (b)~the transformation of these values must be monotonically increasing.
For simplicity of treatment, we shall focus on the case where
monotonicity holds in a strict sense, although this is not logically compelling;
on this point, see also the final paragraph of Section~\ref{s:transform}.
Combining remarks (a) and (b), we are led to consider transformations
of the type $q_{k}=Q(u_{k})$ where $Q$ is a continuous quantile function,
that is, the inverse of a continuous distribution function $P=Q\inv$. 
Hence the $q_{k}$'s represent the quantiles of $P$
at  the levels $k/(K+1)$, for $k=1,\ldots, K$.

From the formal viewpoint, we have not yet made any progress,
since the $q_{k}$'s simply represent an arbitrary 
sequence of numbers, with only the condition of monotonicity.
Their interpretation as quantiles, however, lends itself 
to the introduction of a flexible parametric family of distributions 
from which to select a suitable member $P$.
We shall sometimes write $q_{k}(\theta)=Q(u_{k}; \theta)$
to emphasize the dependence on the parameter vector $\theta\in\Real^p$
which identifies a given distribution in the parametric family.

It is widely recognized that, at least in the univariate context 
like the present one, 
flexible parametric families can be used to approximate closely 
an arbitrary continuous distribution, except possibly very peculiar situations.
Some numerical evidence in this direction has been provided by
\cite{solo:step:1978} using the Pearson system of curves, 
but the underlying idea applies more generally.
For instance, \cite{hoaglin:1984} illustrates how the $g$-and-$h$ family
can satisfactorily approximate a given target distribution, such as
$\chi^{2}_{6}$ and Cauchy; the $g$-and-$h$ family will be recalled later.
We shall return to the use of flexible parametric families in a short while.

Before proceeding further,  
we summarize the essence of the proposal, which is as follows: 
(i)~adopt a flexible parametric family of continuous distributions;
% typically with unbounded range;
(ii)~select its member $P$ whose quantile values  $q_{k}$'s 
are most effective at fitting the data under consideration,
and denote these quantiles by $q_{k}(\hat\theta)$;
(iii)~optionally, these $q_{k}(\hat\theta)$ are transformed on the original scale to
produce new working scores, for instance adopting $(K{+}1)\,q_{k}(\hat\theta)$ 
as the final scores.
In next subsection, we discuss the more operational side of steps (i) and~(ii).

%----------------
\subsection{On the choice of the transformation} \label{s:transform}

To start with, the methodology requires to select a flexible parametric 
family of univariate continuous distributions. We discuss some guidelines 
for its choice. 

Unless the specific problem under consideration incorporates some features 
which indicate differently, it is appropriate to work with distributions
having support on the entire real line, to avoid unnecessary restrictions
on the $q_{k}$'s. Typically, such distributions include
parameters for the regulation of location, scale and shape, with shape
often regulated by more than one parameter. 
For our purposes, however, location and scale can be fixed at 0 and 1, say, 
since their effect can be subsumed under the other components 
of the (generalized) regression model.
Hence, we are only interested in regulating the shape parameters, 
of which there are typically two.

Even with these considerations, the choice is among a vast set of options.
Consider that the fitting process involves the computation of 
$Q(u_{k}; \theta)$ for all values of $k$ and for many candidate $\theta$'s. 
From the computational viewpoint, it is then advantageous to work
with a family of distributions whose quantile function can be computed
efficiently, ideally via an explicit expression.

Along this line of thinking, a  convenient setting is represented by
families obtained from transformations of the normal distribution.
A classical construction of this type is the
$S_{U}$ distribution introduced by \cite{johnson:1949}; 
the main ingredients are summarized by 
\citet[Chapter 12, \S\,4.3]{john:kotz:bala:1994} under the heading
`Transformed Distributions'.
The essence of the construction is as follows: 
start with a standard normal random variable $Z$ and 
define $\tilde Z=(Z-\gamma)/\delta$ for given real parameters 
$(\gamma, \delta)$, where $\delta$ is positive; 
then  an instance $Y$ of the $S_{U}$ family 
(without inclusion of location and scale parameters) 
is generated  by the transformation
\begin{equation}
  Y = \sinh \tilde Z 
    = \half \left(e^{\tilde Z} - e^{-\tilde Z}\right) .
  \label{e:SU}
\end{equation}
As the parameters $(\gamma, \delta)$ span their admissible range,
the $S_{U}$ family exhibits a high degree of flexibility.
In particular, the measures of skewness and kurtosis vary considerably. 
If $\gamma=0$, the density of $Y$ is symmetric about 0.

Another construction of similar logic is represented by the Tukey 
$g$-and-$h$ distribution, discussed in detail by \cite{hoaglin:1984}. 
Also in this case, we start from the random variable $Z$ as before, 
but its transformation  takes now the form 
\begin{equation}
 Y =  \cases{ %\begin{cases}
       Z\:e^{h\,Z^{2}/2}         & \hbox{if~} g=0,     \cr
       g\inv\left(e^{g\,Z}-1\right) e^{h\,Z^{2}/2} 
                                & \hbox{otherwise} 
  } %\end{cases}                                     
  \label{e:gh}
\end{equation}
for arbitrary parameters $g$ and $h$, provided $h\ge0$.  
Also for the $g$-and-$h$ distribution, the measures of 
skewness and kurtosis span a considerable range.
If $g=0$, the density of $Y$ is symmetric about 0.
If both $g=0$ and $h=0$,  \ref{e:gh} reduces to $Y=Z$.

Yet another construction is the  $\mathrm{sinh}$--$\mathrm{arcsinh}$ 
distribution introduced more recently by \cite{jone:pews:2009}.
This is based on the transformation
\begin{equation}
  Y =  \sinh\left(\frac{\mathrm{arcsinh}\,Z + \eps}{\delta}\right)
  \label{e:sinh-arcsinh}
\end{equation}
where $\eps$ is a real-valued parameter which regulates skewness, 
with sign in agreement with the one of $\eps$, and
$\delta$ is a positive parameter which regulates tailweight.
When $\eps=0$ and $\delta=1$, \ref{e:sinh-arcsinh} reduces to $Y=Z$.
A convenient feature of \ref{e:sinh-arcsinh} is that it can
generate both lighter tails than the normal (if $\delta>1$)
and heavier ones  (if $\delta<1$).

While the $S_{U}$, the $g$-and-$h$ and the $\mathrm{sinh}$--$\mathrm{arcsinh}$ 
families  appear to be suitable for our purposes, 
there is no compelling reason for their choice. 
In specific problems, other families of distributions
could possibly be preferable.

A reviewer of this paper has suggested considering the Beta family
because of its high flexibility as the parameters vary. 
Use of the Beta distribution had indeed been explored in an early stage
of this project, but the outcome was disappointing.
The reason of the unsuccess seemed to be linked to the bounded
support of the distribution, while the extemal classes of the factors 
typically call for wide ranges of the quantiles.
The issue of restricted range of the distribution cannot be
overcome by considering an arbitrary range $(a, b)$, say,
instead of $(0,1)$, regarding $(a,b)$ as additional parameters,
for the same reason that has excluded the location 
and scale parameters for the distributions mentioned above.

Another interesting suggestion was to allow that some of the quantiles 
of $Q$ coincide, so that  two or more factor levels could effectively 
collapse into one, a situation which may be appropriate in some cases.
Technically, this extension means removing the condition that $Q$ is strictly
monotonic. To achieve this situation, we can consider a transformation 
distribution  obtained as a mixture of a point mass and a continuous distribution.
This formulation is interesting, and deserves to be examined 
in future developments, but for the present presentation we 
focus on the simpler situation of strictly monotonic transformations,
corresponding to preservation of distinct levels of the factor.

%---------------------------------------------------------------------
\section{Numerical illustrations} \label{s:examples}

%-----------------------

\subsection{Oesophageal cancer data} \label{s:example1}

To demonstrate the working of the method in a concrete situation,
we consider a set of data referring to a case-control study of
oesophageal cancer reported by \cite{bres:day:1980}, 
and available in the \Rlang\ computing environment as dataset \texttt{esoph}.
The aim of the exercise is to model the number of cases and controls
using three numerical explanatory variables which have been grouped 
into categories, hence yielding three ordered factors.
Specifically, there are six groups for age (factor \texttt{agegp}), 
four groups for alcohol consumption (\texttt{alcgp}), and 
four groups for tobacco consumption (\texttt{tobgp}).

Logistic regression models  have been fitted to the number 
of cases and controls,  via the \Rlang\ command \texttt{glm},
with a linear predictor built from the above-described explanatory variables.
After fitting an initial model with maximal degrees of the 
orthogonal polynomials representing the ordered factors, 
the non-significant components have been removed, 
leading to the model summarized in Table~\ref{t:m3a}.
Here and in the following, the suffixes `.L', `.Q' and `.C' 
denote the linear, quadratic and cubic terms of a factor.
Although the quadratic component of \texttt{alcgp} is not significant,
it has been retained because the cubic component is, 
in order to follow a hierarchical building scheme.

% latex table generated in R 4.2.1 by xtable 1.8-4 package
% Fri Mar 10 18:38:29 2023
\begin{table}[ht]
\caption{\sl Oesophageal cancer data: summary ingredients of the logistic regression 
using orthogonal polynomials for representing the ordered factors effects.}
\label{t:m3a}
\vspace{1ex}
\centering
\begin{tabular}{rrrrr}
  \hline
 & Estimate & Std.\,Error & z value & Pr($>$$|$z$|$) \\ 
  \hline \tt
 (Intercept) & -1.154 & 0.170 & -6.79 & 0.000 \\ 
  agegp.L & 3.706 & 0.433 & 8.55 & 0.000 \\ 
  agegp.Q & -1.481 & 0.398 & -3.72 & 0.000 \\ 
  tobgp.L & 0.966 & 0.215 & 4.50 & 0.000 \\ 
  alcgp.L & 2.505 & 0.258 & 9.73 & 0.000 \\ 
  alcgp.Q & 0.082 & 0.220 & 0.37 & 0.708 \\ 
  alcgp.C & 0.398 & 0.181 & 2.20 & 0.028 \\ 
  \hline
  \multicolumn{5}{c}{   Residual deviance: 88.215 }
\end{tabular}\par
\end{table}

For this initial illustration, we focus on the \texttt{alcgp} factor, 
and apply the method introduced in Section~\ref{s:method} only to \texttt{alcgp}.
The linear, quadratic and cubic effects
of the factor \texttt{alcgp} in Table~\ref{t:m3a} are now replaced
by a single explanatory variable denoted \texttt{alcgp.scores}.
Two variants have been developed:
one using the $S_{U}$ and the other using the $g$-and-$h$ distribution. 
Numerical minimization of the residual deviance computed at different
values of the transformation parameters has produced the values
$(\gamma, \delta)=(-0.025, 0.395)$ for $S_{U}$ 
and $(g,h)=(0.116, 1.85)$ for $g$-and-$h$.
The outcomes of the corresponding logistic regressions are reported 
in Table~\ref{t:m4}.

%
%--------------------------------------------
% latex table generated in R 4.2.1 by xtable 1.8-4 package
% Fri Mar 10 19:21:13 2023
\begin{table}[th]
\caption{\sl Oesophageal cancer data: summary ingredients of the logistic
regression where the \texttt{alcgp} factor is expressed via a constructed 
set of scores.
The first sub-table employs the $S_{U}$ distribution; the second sub-table 
uses a $g$-and-$h$ distribution. }
\label{t:m4}
\centering
\vspace{2ex}
\makebox[11em]{$S_{U}$ distribution \hfill } 
\begin{tabular}{rrrrr}
  \hline
 & Estimate & Std.\,Error & z value & Pr($>$$|$z$|$) \\ 
  \hline
  \tt
 (Intercept) & -1.228 & 0.165 & -7.44 & 0.000 \\ 
  agegp.L & 3.706 & 0.432 & 8.58 & 0.000 \\ 
  agegp.Q & -1.481 & 0.394 & -3.76 & 0.000 \\ 
  tobgp.L & 0.966 & 0.215 & 4.50 & 0.000 \\ 
  alcgp.score & 0.427 & 0.042 & 10.06 & 0.000 \\ 
  \hline
  \multicolumn{5}{c}{Residual deviance: 88.215}
\end{tabular} 
\par
%----
%\begin{minipage}{0.4\hsize}
%??
%\end{minipage}
% latex table generated in R 4.2.1 by xtable 1.8-4 package
% Sat Mar 11 09:14:13 2023
\vspace{2ex}
\makebox[11em]{ $g$-and-$h$ distribution\hfill}
\centering
\begin{tabular}{rrrrr}
  \hline
 & Estimate & Std.\,Error & z value & Pr($>$$|$z$|$) \\ 
  \hline
  \tt
(Intercept) & -1.200 & 0.165 & -7.25 & 0.000 \\ 
  agegp.L & 3.706 & 0.432 & 8.58 & 0.000 \\ 
  agegp.Q & -1.481 & 0.394 & -3.76 & 0.000 \\ 
  tobgp.L & 0.966 & 0.215 & 4.50 & 0.000 \\ 
  alcgp.score & 1.089 & 0.108 & 10.06 & 0.000 \\ 
   \hline
  \multicolumn{5}{c}{Residual deviance: 88.215} 
\end{tabular} 
\end{table}
%--------------------------------------------

Using the residual deviance as the target criterion for the selection
of the transformation parameters seems quite a natural choice, since
it coincides with the criterion used for fitting the logistic regression model of interest.
In case the fitting criterion adopted for the original logistic regression 
model was different, such as some form of regularized log-likelihood, 
the same criterion could equally be adopted for selecting the 
transformation parameters.
However, while this logic `by analogy' seems the more natural one, 
alternative procedures are not ruled out, in principle.

A noticeable feature emerging from the comparison of the
entries in Tables~\ref{t:m3a} and~\ref{t:m4} is the almost exact
coincidence of several components, notably the residual deviance and 
the estimates and standard errors of \texttt{agegp.L},
\texttt{agegp.Q} and \texttt{tobgp.L}.
Therefore, the modified handling of \texttt{alcgp} 
has not altered the outcomes associated to the other components
of the logistic model, which is a welcome feature.

The differences are in the intercept term and, for the two
subtables of Table~\ref{t:m4}, in the estimates and standard errors
of \texttt{alcgp.score}.
These differences merely reflect the different numeric ranges
of the two sets of scores, but their actual working is the same,
as we discuss next.
First of all, the $z$-values on the two lines of \texttt{alcgp.score} 
coincide, and so the same holds for the observed  significance level.

\begin{figure}
\includegraphics[width=0.49\hsize]{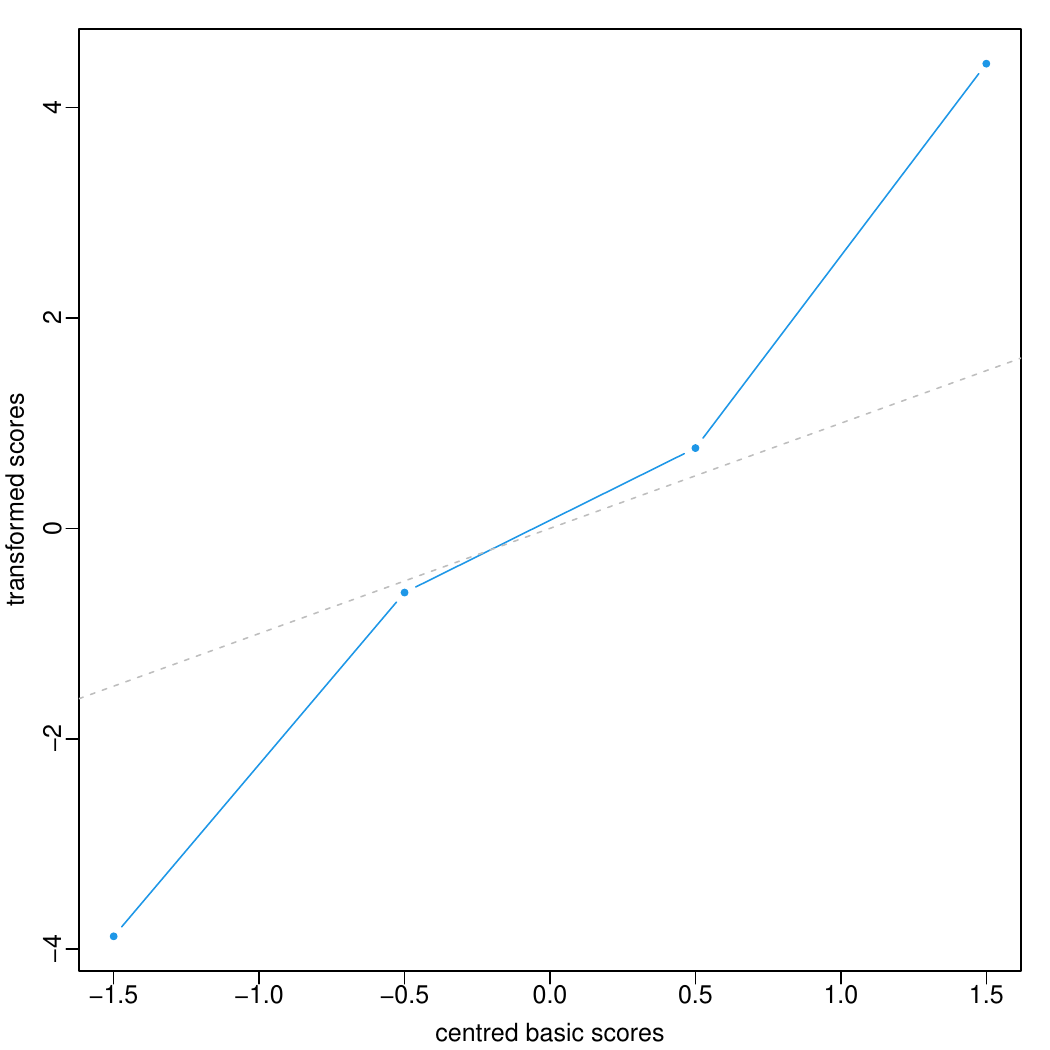}\hfill
\includegraphics[width=0.49\hsize]{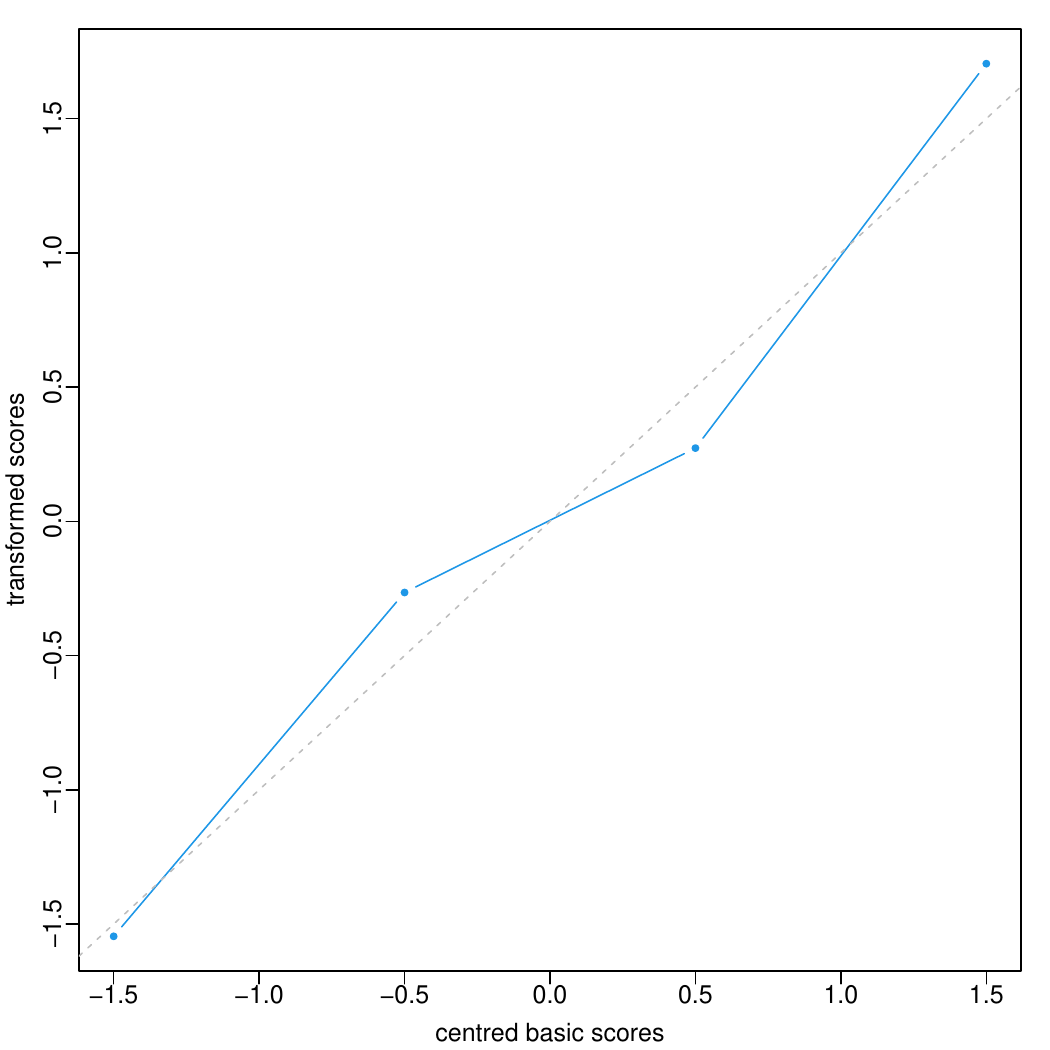} 
\caption{\sl Oesophageal cancer data: new scores versus equally spaced scores
  using the $S_{U}$ (left panel) and $g$-and-$h$ distribution (right panel);
  the grey dashed line is the identity.}
\label{f:newscores}
\end{figure}

Additional insight can be obtained from close consideration of the scores
values, visualized in the plots of Figure~\ref{f:newscores}.
For the $S_{U}$ family, the parameter estimates are $(\hat\gamma, \hat\delta)
\approx(-0.025,  0.395)$, leading to score values $(-3.88, -0.61, 0.76, 4.42)$,
using transformation \ref{e:SU} applied to $u_{1},\dots, u_{4}$.
These scores are plotted on the left panel of Figure~\ref{f:newscores}
versus the basic scores $(1,2,3,4)$ shifted so that they are centred on $0$,
for easy comparison. It is clear, both numerically and graphically, that
the new scores have the values of the extremal classes more stretched out.
This elongation is almost symmetric on the two sides, and it would 
be conceivable to fit a constrained $S_{U}$ distribution with $\gamma=0$.

The indication from the right panel of Figure~\ref{f:newscores} are 
qualitatively very much the same for the $g$-and-$h$ distribution; 
only the numeric range of the new scores is scaled down, 
namely $(-1.55, -0.26,  0.27, 1.70)$.
The ratio of the $S_{U}$-associated scores are about 2.5 times larger
than those associated to the $g$-and-$h$ distribution and,
correspondingly, the \texttt{alcgp.score} estimates $0.427$ and $1.089$ are 
reversely related.
  
Given the essential coincidence of the numerical outcomes reported in
Tables~\ref{t:m3a} and~\ref{t:m4}, consideration about the relative
merits of the classical and the new methodology rests on the more
qualitative side, especially interpretability of the outcome.
In this respect, it seems advantageous to be provided with a set of
scores which quantify in a simple way how the four categories of
\texttt{alcgp} are actually spaced.  For further simplification,
especially for convenience of non-specialist readers, 
one could  linearly transform the scores to more easily readable ones. 
For instance, we could transform the scores so that the first two of them 
are $(1,2)$, which leads to the values $(1, 2, 2.42, 3.54)$ 
for both the $S_{U}$ and the $g$-and-$h$ variant.

We have not yet mentioned a method for constructing scores 
which has a long tradition  in statistical practice, 
namely the use of mid-values of the \texttt{alcgp} classes.
This method is has the advantage of simplicity, 
but it runs into difficulty when the class in unbounded.
In the example under consideration, the upper class of \texttt{alcgp} 
is $[120, \infty)$ and the selection of a sensible `mid-value' introduces 
another element of subjective choice. In fact, it is quite common that 
at least one of the extremal classes of an order factored constructed
by grouping of continuous values is unbounded.
Another limitation of this option is that it is feasible only for
factors obtained by grouping of an underlying continuous variable;
it is not feasible with a different genesis of the factor, 
such as in the case considered in the next numerical illustration.

%-----------------------
\subsection{Diamond pricing data} \label{example2}

Various packages within the \Rlang\ environment,  such as \texttt{ggplot2}, 
include the dataset  \texttt{diamonds}  which comprises 10 characteristics of 54{,}940 
round cut diamonds. The main determinants of the price are four 
classical variables of interest, often collectively denoted as ``the 4C's'', 
which are as follows:\par
\begin{tabular}{rl}
\texttt{carat}  & denotes weight (1 carat equals $0.2$ grams);  \\
\texttt{clarity} & indicates how clear the diamond is,  \\
                & with   levels I1 (worst), SI2, SI1, VS2, VS1, VVS2, VVS1, IF (best);\\
\texttt{color} & the diamond colour,  from D (best) to J (worst);\\
\texttt{cut} & indicates the quality of the cut, 
               with levels Fair, Good, Very Good, Premium, Ideal. 
\end{tabular}
\vspace{1ex}
\par\noindent
Among these, only \texttt{carat} is of numeric type;
all the others are of ordinal type.
The purpose of the exercise is to find a simple rule for predicting
the monetary value (in USD)  from the 4C's variables.
Given the large number of cases in the dataset and the purely illustrative
aim of the present demonstration, we examined  a small subset of the data, 
namely those with row number $1, 101, 201, \ldots$.

The instinctive starting point is a linear model with response variable 
\texttt{price} and  explanatory variables \texttt{carat}, \texttt{clarity},
\texttt{color} and  \texttt{cut}. 
However, graphical exploration of the data  exhibited noticeable curvature
of the response variable as well as increasing dispersion as \texttt{carat} increases.
Use of the Box-Cox transformation indicated $0.436$ as the appropriate power
transformation for linearizing the relationship, although this does not quite
stabilizes the residual dispersion. 
For simplicity, the power value $0.436$ has been approximated to $0.5$
even if this value is slightly outside the pertaining confidence interval,
hence using \texttt{sqrt(price)} as the response variable.
The ordered factors have been initially included using orthogonal polynomials
at the maximal degree allowed by their number of levels, which are $8, 7, 5$,
respectively.
Observations numbered $(518, 519, 523)$  resulted
to be clear outliers with high leverage; after their removal, there were $537$
observations left.
In addition, several higher order components of the polynomials could be 
removed, leading to the model summarized in Table~\ref{t:m06b}. 
 
\begin{table}[ht]
\caption{\sl Diamond pricing data: summary ingredients of the linear model 
for response variable \texttt{sqrt(price)}
using orthogonal polynomials for expressing the ordered factors effects.}
\label{t:m06b}
\vspace{1ex}
\centering
\begin{tabular}{rrrrr}
  \hline
   & Estimate & Std.\,Error & z value & Pr($>$$|$z$|$)  \\ 
  \hline \tt
  (Intercept) & 1.194 & 0.75 & 1.60 & 0.11 \\ 
  carat & 65.317 & 0.71 & 92.31 & 0.00 \\ 
  clarity.L & 24.154 & 1.58 & 15.25 & 0.00 \\ 
  clarity.Q & -11.701 & 1.24 & -9.42 & 0.00 \\ 
  clarity.C & 3.610 & 1.24 & 2.90 & 0.00 \\ 
  color.L & -13.271 & 0.96 & -13.76 & 0.00 \\ 
  color.Q & -1.907 & 0.89 & -2.13 & 0.03 \\ 
  color.C & 1.979 & 0.85 & 2.33 & 0.02 \\ 
  color\verb|^|4 & 3.369 & 0.78 & 4.30 & 0.00 \\ 
  cut.L & 1.882 & 0.85 & 2.21 & 0.03 \\ 
   \hline
  \multicolumn{5}{c}{Residual std.\ deviation: 6.74 on 527 degrees of freedom }
\end{tabular}
\end{table}

For this numerical illustration, we apply the proposed method to two factors
simultaneously, namely  \texttt{clarity} and \texttt{color}, making use of
the $g$-and-$h$ and the $\mathrm{sinh}$--$\mathrm{arcsinh}$ distribution.
Since both distributions feature two parameters, the numerical optimization
for minimizing the residual variance involves a four-dimensional search
in both cases. 
Similarly to the illustration of Section~\ref{s:example1}, the target criterion 
for choosing the trasformation parameters agrees with the criterion for fitting 
the original model, that is, the residual variance in case of a linear model;
again, penalized variants of the residual variance are possible.
The outcome of the fitting process is summarized in Table~\ref{t:m06b.score}.
We do not report the analogous values using the $S_{U}$ distribution,
but they were qualitatively similar to those for the other two distributions.

\begin{table}[ht]
\caption{\sl Diamond pricing data: summary ingredients of the linear model 
for response variable \texttt{sqrt(price)}
when the factors \texttt{clarity} and \texttt{color} are expressed 
via constructed  sets of scores.
The first sub-table employs the  $g$-and-$h$ distribution; 
the second sub-table refers to the $\mathrm{sinh}$--$\mathrm{arcsinh}$ distribution. }
\label{t:m06b.score}
\centering
\vspace{2ex} %----------------
\makebox[11em]{g-and-h distribution \hfill } 
\begin{tabular}{rrrrr}
  \hline
 & Estimate & Std.\,Error & t value & Pr($>$$|$t$|$) \\ 
  \hline \tt
  (Intercept) & 6.448 & 0.67 & 9.68 & 0.00 \\ 
  carat & 65.001 & 0.71 & 91.03 & 0.00 \\ 
  clarity.score & 8.618 & 0.47 & 18.52 & 0.00 \\ 
  color.score & -6.811 & 0.49 & -13.98 & 0.00 \\ 
  cut.L & 2.003 & 0.87 & 2.31 & 0.02 \\ 
  \hline
  \multicolumn{5}{c}{Residual std.\  deviation: 6.90 on 532 degrees of freedom}
\end{tabular}
\par
\vspace{2ex} %----------------
\makebox[11em]{$\mathrm{sinh}$--$\mathrm{arcsinh}$ distribution \hfil} 
\begin{tabular}{rrrrr}
  \hline
 & Estimate & Std.\,Error & t value & Pr($>$$|$t$|$) \\ 
  \hline \tt
  (Intercept) & 23.105 & 1.00 & 23.05 & 0.00 \\ 
  carat & 65.087 & 0.71 & 91.72 & 0.00 \\ 
  clarity.score & 0.000 & 0.00 & 18.72 & 0.00 \\ 
  color.score & -0.072 & 0.01 & -14.26 & 0.00 \\ 
  cut.L & 2.028 & 0.86 & 2.35 & 0.02 \\ 
  \hline
  \multicolumn{5}{c}{Residual std.\ deviation: 6.85 on 532 degrees of freedom}  
\end{tabular}
\end{table}

Taking the residual standard deviation as the reference ingredient
in Table~\ref{t:m06b.score},
we see a modest increase with respect to the value, $6.74$, of Table~\ref{t:m06b}.
On the other hand, there is a substantial simplification of the model complexity,
which may more than compensate the limited increase of residual standard deviation.
A formal comparison between the model of Table~\ref{t:m06b} and those of
Table~\ref{t:m06b.score} is not feasible by standard procedures, because 
the models are non-nested, and in addition because of the qualitative difficulty
of making allowance for the parameters of the transformation distributions;
on a broadly connected point, see also the discussion of Section~\ref{s:discussion}.

A specific note is due for the sub-table of the 
$\mathrm{sinh}$-$\mathrm{arcsinh}$ distribution. 
The very small absolute values  of some estimates reflect the very large
size of the constructed scores, especially so for \texttt{clarity} whose
scores range up to $10^9$. 
Correspondingly, the \texttt{clarity.score} estimate appearing as $0.000$ in 
Table~\ref{t:m06b.score} becomes $7.83{\times}10^{-9}$ with standard error
$4.18{\times} 10^{-10}$  when written  in scientific notation. 
In practical work, these scores could simply be scaled down to 
more manageable values, with matching increase of the estimates 
and standard errors. However, for the present illustration, 
we have preferred to report the outcome in its original form.  
 
%---------------------------------------------------------------------
\section{Some miscellaneous remarks} \label{s:discussion}

As stated in the introductory section, the aim of this proposal is
to retain an adequate fitting, as compared to the currently standard
technique based on the polynomial representation of the factor,
while providing a plainer interpretation of the resulting model.
Note that the methodology does not attempt to improve  on the 
polynomial-based fitting  when one is ready to set a sufficiently high degree 
of the polynomial,  since this scheme will eventually involve as many parameters 
as the number of factor levels. On the other hand, a high-degree polynomial
model goes against the idea of a parsimonious model, and 
it is not desirable from an interpretational viewpoint, 
while these aspects constitute key motivations for the present proposal.

Clearly, the proposed method can be employed in a variety of situations,
not only linear or generalized linear models. 
The method is suitable for any formulation where the response variable 
is regulated, possibly through some transformation, by some predictor 
which incorporates the ordered factors.
For instance, proportional hazard models for the analysis of 
survival data represent another possibility.
In this case, the natural criterion for choosing the transformation
parameter is maximization of the partial log-likelihood, possibly
in some penalized form.

In the numerical cases which have been examined, 
including those which have not been reported here,
adequate linearization of the factor effect could always be achieved 
with some suitably chosen parameters of the transformation.
It could possibly happen that in harder problems such a linearization 
may be not attainable; in these cases, one could consider introducing 
a parametric family of distributions with increased flexibility, 
involving more than two shape parameters. 
For instance, one could consider the 
generalized hyperbolic distribution, which allows higher flexibility, 
although at the cost of an higher computational effort and
often increased difficulties at the fitting stage.

A possible remark concerning the interpretation of the outcome
summarized in Table~\ref{t:m4} and \ref{t:m06b.score} is that the 
\texttt{<factor>.score} coefficients vary with the parameters 
of the transformation distribution and we should consider whether
this variability should be taken into account.
More explicitly, the question is whether the evaluation of 
the variance of the parameter of interest, namely the coefficients of 
\texttt{alcp.score}, \texttt{clarity.score} and \texttt{color.score},
should include its inflation due to the estimation step 
of the transformation parameters.
This sort of consideration is similar to one raised for the 
Box-Cox transformation in other situations.
Recall that, while the Box-Cox transformation had been introduced 
for the response variabile, it has been used also 
for transforming explanatory variables. 
The issue is quite broad and beyond the scope of the present note,
but we mention at least the situation examined by \cite{siqu:tayl:1999}
because of the similarity with the one of Section~\ref{s:example1},
since they also deal with a logistic model, the difference being that they 
consider a Box-Cox transformation of a continuous explanatory variable. 
One simple way of addressing the issue of variance inflation here is similar
to the one denoted `conditional' by \cite{siqu:tayl:1999}, but its
essence can be linked to the argumentation of \cite{box:cox:1982}.
Under this view, the interpretation of the regression coefficient must refer
to a fixed transformation, hence to fixed transformation parameters, 
since the assessment across different transformations would imply
combining values on different and incomparable scales, preventing any  
sound scientific interpretation.

The proposed methodology has been implemented in an \Rlang\ package
freely available at \url{https://CRAN.R-project.org/package=smof} 
whose documentation also shows how to reproduce the numerical
illustrations of Section~\ref{s:examples}.
 
\vspace{3ex} 
\subsection*{Acknowledgements} 
I am grateful to Sandy Weisberg and Alan Agresti for a number of 
insightful remarks on a preliminary sketch of this proposal, 
leading to a substantially improved presentation. 
Additional constructive comments have been provided by an associate editor
and four reviewers of the submitted paper, to whom I also express my gratitude
for stimulating an improved exposition.

%------------------------------------------------------------------------------
\clearpage
%\small

\end{document}